\documentstyle[editedbook,epsfig,psfig,epsf]{mq}
\def\G1915{GRS $1915$+$105$}
\def\X1550{XTE J$1550$--$564$}
\def\J1655{GRO J$1655$--$40$}

\def\cm2{cm$^2$ }
\def\se1{s$^{-1}$ }

\begin{opening}
\title{Evolution of a low frequency QPO during the 2000 outburst of XTE J1550--564}
\author{J. Rodriguez$^1$, S. Corbel$^{1,2}$, E. Kalemci$^{3}$, J.A. Tomsick$^{3}$, M. Tagger$^1$}
\institute{$^1$ CE Saclay, DSM/DAPNIA/Service d'Astrophysique (CNRS URA 2052), F-91191  Gif sur Yvette France.\\
$^2$ Universit\'e Paris 7, 5 place Jussieu, F-75005 Paris, France.\\
$^3$ Center for Astrophysics and Space Science, UCSD, San Diego, USA}
\end{opening}

\runningtitle{XTE J1550--564 : low frequency QPO}
\runningauthor{Rodriguez et al.}

\begin{document}
\vspace{-0.5cm}
\begin{abstract}
{\small We follow the evolution of a low frequency QPO during the 2000 outburst
 of the microquasar XTE J1550--564, which was found to be present in the PCA 
energy range (2--65 keV) in 19 of 43 observations. The frequency of the QPO 
varies from 0.1 Hz to 6 Hz, and appears to follow the evolution of the soft
 X-ray flux. If we assume the soft X-rays represent the behavior of an 
accretion disk, the relation indicates that this low frequency QPO is linked 
to the accretion disk. We show that the non-trivial relation between the QPO 
frequency and the soft flux may be as expected from the Accretion Ejection
 Instability (AEI), when the disk approaches its last stable orbit. 
Furthermore, the energy dependence of the QPO may indicate the presence of a 
hot spot rotating in the disk as predicted by the AEI.}
\end{abstract}

\section{AEI and QPO in microquasars}
Low frequency Quasi Periodic Oscillations (LFQPO) are commonly observed in 
microquasars when a strong power law component is detected in 
their X-ray energy spectra, with a typical frequency of 0.1--10 Hz. 
During the Low Hard State (LHS) they have high amplitude ($\sim 15\%$ rms), 
and during the Intermediate/Very High State (IS/VHS) they have a moderate 
amplitude ($\sim 5\%$ rms). Their presence can be explained in the 
context of the Accretion Ejection Instability (AEI, \cite{Tagger99}), which 
 occurs in the innermost region of an accretion disk threaded by vertical 
magnetic field lines. The instability manifests as a spiral density wave 
(a hot point) rotating at  $10-30 \%$ of the Keplerian frequency at the
 inner edge of the disk, producing the modulation detected as a LFQPO. One 
would then expect the frequency $\nu$ of the QPO to vary as 
$\nu \propto r^{-3/2}$, r being the inner radius of the disk. We have shown 
\cite{Varniere02} that due to general relativistic 
effects this relation is modified whenever the disk is close to its last 
stable orbit (LSO). This could explain the behavior we observed in the 
case of \J1655, where the relation between the QPO frequency and the disk 
inner radius was inverted compared to \G1915 \cite{Rodriguez02a}, whereas the 
energy dependence of the QPO amplitude in \G1915 may suggest the presence 
of a hot point in the disk \cite{Rodriguez02b}.

\section{Spectral Overview of the 2000 outburst}
\begin{figure}[htbp]
\centering
\epsfig{file=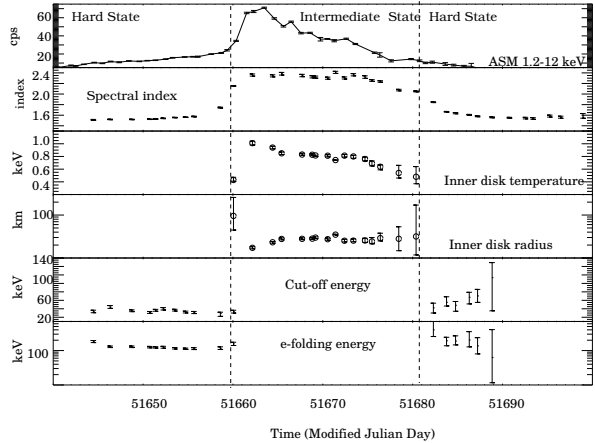,width=6cm,angle=-90}
\vspace*{-0.2cm}
\caption{Overview of the spectral evolution of XTE J1550--564 during the 2000
 outburst. Vertical lines indicate the dates of state transitions.}
\label{fig:overview}
\end{figure}

\X1550 is a microquasar \cite{Hannikainen01} hosting a black hole of 
$M=10.5 \pm 1.5$ $ M_\odot$ \cite{Orosz02}, located at  $\sim 5.3-5.9$ kpc 
\cite{Orosz02}. On 2000 April 6 (MJD 51640), \X1550 became active 
\cite{Smith00} undergoing an episode of outburst. From MJD 51644--51690 the 
source was monitored with RXTE. We fitted the PCA+HEXTE spectra between 
$3$ and $150$ keV with a model consisting of interstellar absorption, a 
smeared Iron edge at 
$\sim 7$ keV, and a power law. A high energy cut-off is needed to fit some  
spectra, and a thermal component is also included in some others. 
The evolution of the spectral parameters is plotted in Fig. \ref{fig:overview}.
Based on the spectral evolution of the source we estimate
 that  \X1550 has transited from a low hard state (LS, photon index $<$2) 
(MJD 51644--51658) into an intermediate state (IS, photon index $>$2, thermal 
component) between  MJD 51658 and MJD 51660. It stayed in 
that state until MJD 51680, and then transited back to a LS at least until
the end of our study \cite{Rodriguezmoriond}.\\
\vspace{-0.8cm}
\vspace{-0.3cm}

\indent During the IS the small values 
of the color radius returned from the fits, and their relative constancy over 
$\sim 10$ days suggest that the inner disk is close to its LSO, in good 
agreement with the detection of HFQPO \cite{Miller01}.The disk 
contributes a fairly constant $2-50$ keV luminosity of $40-50 \%$ 
during most of the IS, and after MJD 51674 the temperature starts 
to decrease and the disk becomes undetectable on MJD 51682. Although the late 
behavior of the inner radius is relatively difficult to constrain given the 
errors on this parameter, the disappearance of the thermal component on 
MJD 51682, and the  observation of a 65 Hz QPO \cite{Kalemci01} favors an 
interpretation where as the inner edge of the disk is moving outward, its 
maximum temperature drops down to low values.

\section{Evolution of the QPO frequency}
\begin{figure}[htbp]
\centering
\epsfig{file=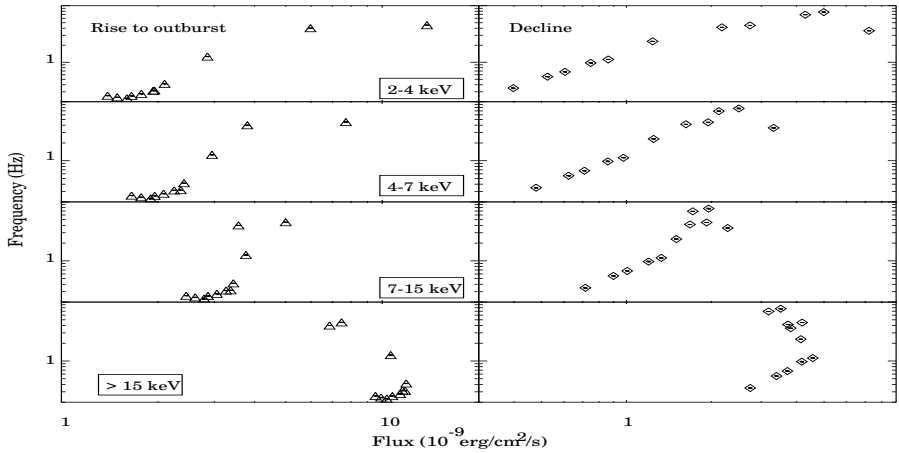,width=12cm,height=6cm}
\vspace*{-0.2cm}
\caption{Evolution of the QPO frequency with the source flux in different energy ranges during both the rise (left) and the decline (right) of the outburst.}
\label{fig:QPOvsflux}
\end{figure}

For both epochs the evolution of the QPO frequency (Fig. \ref{fig:QPOvsflux}) 
correlates better with 
the $2-4$ and $4-7$ keV band fluxes. If the evolution 
of the soft flux follows the behavior of the disk radius, as suggested by our 
spectral analysis, then the energy dependent frequency behavior favors an 
interpretation where the LFQPO is somehow related  to the evolution of the 
disk. During the rising part, the energy spectra of \X1550 are dominated by 
the power law component. If that 
component characterizes an inverse Compton effect of the soft disk photons as 
usually assumed, the rise to outburst may indicate that the disk is approaching
 the black hole during the initial low state. Hence, although the 
frequency--flux relation is not absolutely linear (left part of Fig. 
\ref{fig:QPOvsflux}), the increase of the frequency with the flux still 
favors an association with the disk. When the disk brightens, in particular 
when it becomes detectable in the spectra (from MJD 51660 through MJD 51680, 
Fig. \ref{fig:overview}), the plot (Fig. \ref{fig:QPOvsflux}, right) becomes 
linear, which is predicted by the AEI. The plateau and the inversion of the 
slope (Fig. \ref{fig:QPOvsflux})
 occur at high soft flux, at times where the disk reaches the highest 
temperatures, and contributes the most to the energy spectra.  If the disk 
is close to the LSO during the IS, and on MJD 51674 when the QPO 
re-appears, the decrease of the QPO frequency after the plateau 
(Fig. \ref{fig:QPOvsflux}) might have the same origin as the frequency-radius 
inversion  of the correlation  observed in the case of \J1655 
 \cite{Rodriguez02a}, and is what one would expect from 
the theoretical predictions of the AEI \cite{Varniere02}: when the 
disk is close to the LSO, the rotation frequency of the spiral/vortex is 
modified due to general relativistic effects. 
As the disk is moving outward (decaying soft luminosity, and no detection 
of the thermal component during the final LS), the frequency 
of QPO decreases as expected  when relativistic effects are negligible. 
\vspace{-0.5cm}
\vspace{-0.3cm}

\section{Energy dependence of the LFQPO}
\begin{figure}[htbp]
\vspace{-0.5cm}
\begin{tabular}{ll}
\hspace{-1.5cm}
\epsfig{file=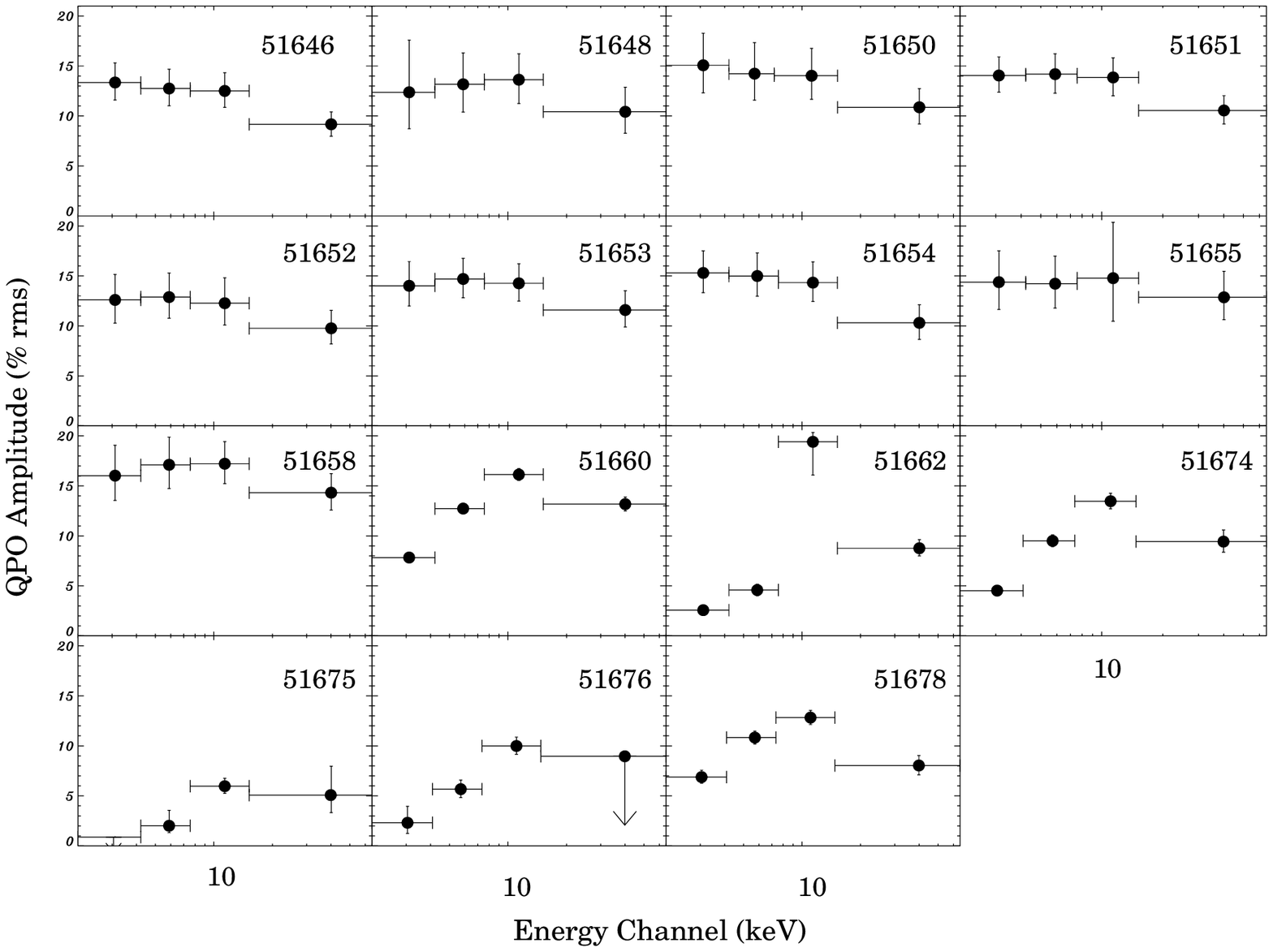,width=8cm,height=5.4cm}&
\epsfig{file=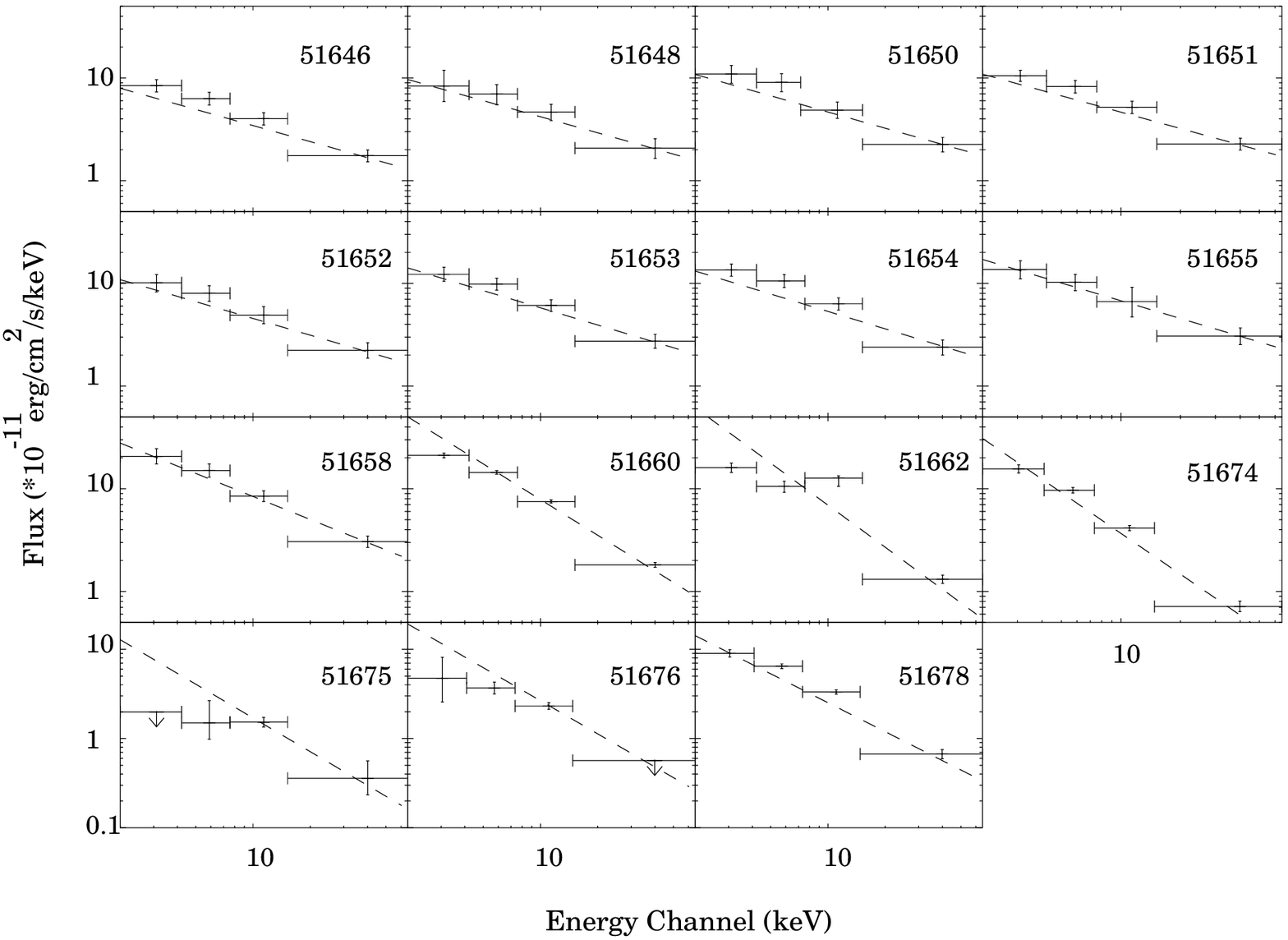,width=8cm,height=5.4cm}\\
\end{tabular}
\vspace*{-0.3cm}
\caption{Energy dependence of the QPO amplitude (left) and flux (right) over
 the outburst. The small number in each panel indicates the date of the observation.}
\label{fig:spectra}
\end{figure}

The energy dependence of the QPO amplitude (Fig. \ref{fig:spectra} left) 
suggests a strong coupling between the Compton medium and the modulation, since
 the amplitude is high (in particular at low energies) when the source 
spectrum shows a strong power law (LS), and vice versa. If we assume the 
power-law flux is modulated and the soft component is un-modulated 
(i.e. the flux from the soft component does not contain the QPO), the 
flattening of the QPO spectrum at low energies (Fig. \ref{fig:spectra}) 
may come from the dilution of the power law by the soft component flux. However
 when we look at the QPO spectra (Fig. \ref{fig:spectra} right), where we 
plotted the modulated flux as a function of the energy channel, and 
superimposed the power law returned from the spectral fits (i.e. with the same 
spectral index), it appears that during the LS there is a slight excess at 
intermediate energies, whereas during the IS there is a flattening in the soft 
X-rays. Our observations may suggest that the QPO originates from an 
independent medium, dynamically linked to the disk. The QPO flux at high 
energies would then simply reflect the inverse Compton reprocessing of 
soft photons on the coronal electrons, while at lower energy this additional 
component would explain the observed spectra.\\
\indent The addition of a 1.3 keV blackbody (not shown) may allow 
a better fit to the spectra, at least during the initial LS 
\cite{Rodriguez02c}. 
This would be consistent with the theoretical predictions of the AEI. In that 
case, the QPO would be the direct signature 
of the instability as it forms shocks (spiral) and warms locally the disk, 
giving birth to a local hot point a bit warmer than the entire disk in its 
average.

\section{Conclusion}
We propose that theoretical predictions of the AEI model match many  
observational constraints, starting with the non trivial evolution of the QPO 
frequency vs. the (soft) flux, and by extension the disk radius. We show that 
a hot point rotating in the disk, which is an expected signature of the AEI, 
is compatible both with the frequency evolution of the flux modulation and the
 energy spectrum of the QPO.

\section*{Acknowledgments}
The authors would like to acknowledge P. Durouchoux and P. Varni\`ere for 
usefull discussions.

\end{document}